\documentclass[sigconf]{acmart}

\renewcommand\footnotetextcopyrightpermission[1]{} 
\usepackage{cleveref}

\usepackage{bm}
\usepackage{bbm}
\usepackage{caption}
\usepackage{subcaption}
\usepackage{booktabs,amsfonts,dcolumn}
\usepackage{mathtools}
\usepackage{wasysym}
\usepackage{hhline}
\usepackage{float}
\usepackage{commath}

 \usepackage{multirow}
 \usepackage[ruled,vlined,linesnumbered]{algorithm2e}
\usepackage{booktabs,siunitx}

\usepackage{stmaryrd}
\usepackage{trimclip}
\usepackage{pifont}
\usepackage{enumitem}
\usepackage{mleftright}
\usepackage{scalerel,stackengine}
\usepackage{bbold}
\usepackage{placeins}

\setcopyright{acmlicensed}
\copyrightyear{2018}
\acmYear{2018}
\acmDOI{XXXXXXX.XXXXXXX}
\acmConference[Conference acronym 'XX]{Make sure to enter the correct
  conference title from your rights confirmation email}{June 03--05,
  2018}{Woodstock, NY}
\acmISBN{978-1-4503-XXXX-X/18/06}

\begin{document}
\title{Reinforcement Learning to Rank Using Coarse-grained Rewards}

\author{Yiteng Tu}
\affiliation{%
    \institution{DCST, Tsinghua University}
    \city{Beijing}
    \country{China}
}
\email{tyt24@mails.tsinghua.edu.cn}

\author{Zhichao Xu}
\affiliation{%
    \institution{University of Utah}
    \city{Salt Lake City}
    \country{United States}
}

\author{Tao Yang}
\affiliation{%
    \institution{Amazon.com Inc}
    \city{Palo Alto}
    \country{United States}
}

\author{Weihang Su}
\affiliation{%
    \institution{DCST, Tsinghua University}
    \city{Beijing}
    \country{China}
}

\author{Yujia Zhou}
\affiliation{%
    \institution{DCST, Tsinghua University}
    \city{Beijing}
    \country{China}
}

\author{Yiqun Liu}
\affiliation{%
    \institution{DCST, Tsinghua University}
    \city{Beijing}
    \country{China}
}

\author{Fen	Lin}
\affiliation{%
    \institution{Tencent}
    \city{Beijing}
    \country{China}
}

\author{Qin	Liu}
\affiliation{%
    \institution{Tencent}
    \city{Beijing}
    \country{China}
}

\author{Qingyao Ai\footnotemark}
\affiliation{%
    \institution{DCST, Tsinghua University}
    \city{Beijing}
    \country{China}
}
\email{aiqy@tsinghua.edu.cn}

\begin{abstract}
Learning to rank (LTR) plays a crucial role in various Information Retrieval (IR) tasks. 
Although supervised LTR methods based on fine-grained relevance labels (e.g., document-level annotations) have achieved significant success, their reliance on costly and potentially biased annotations limits scalability and alignment with realistic goals.
In contrast, coarse-grained feedback signals, such as duration time and session-level engagement, are more accessible and affordable. 
Reinforcement Learning (RL) offers a promising framework to directly optimize these objectives using reward signals, but most existing Reinforcement Learning to Rank (RLTR) approaches suffer from high variance and low sample efficiency. 
Motivated by recent advances in large language models (LLMs), we re-examine the problem of RLTR with coarse-grained rewards and propose new RLTR methods based on widely used RL algorithms for LLMs. 
We systematically compare supervised learning and RL-based methods across various model architectures and coarse-grained reward functions on large-scale LTR benchmarks.
Experimental results demonstrate that advanced RL methods can directly learn from coarse-grained rewards and outperform strong supervised learning baselines even with fine-grained labels. 
This shows the great potential of RLTR for metric-agnostic ranking optimization. 
We release our code at the URL\footnote{\url{https://github.com/StibiumT16/RLTR}}.

\end{abstract}


\maketitle

\section{Introduction}
Ranking is indispensable across multiple Information Retrieval (IR) tasks, including web search~\cite{croft2010search, schutze2008introduction} and recommender systems~\cite{xu2020commerce,zeng2021zero,chen2019top,karatzoglou2013learning}. 
Learning to rank (LTR) typically applies machine learning techniques for ranking documents and items (i.e., ranking candidates)~\cite{wei2017reinforcement,oosterhuis2018ranking,yang2021maximizing,chen2019top,zhao2015multimedia,karatzoglou2013learning,santos2013learning}. 
One popular approach is supervised learning~\cite{burges2010ranknet,cakir2019deep} that applies document-level or item-level relevance annotations to construct and optimize ranking metrics~\cite{oosterhuis2018ranking,jagerman2022optimizing,wang2018lambdaloss} such as Normalized Discounted Cumulative Gain (NDCG)~\cite {jarvelin2002cumulated}. 
While effective, a major drawback of supervised learning is the dependence on fine-grained labels, i.e., explicit relevance judgments for each ranking candidate, to construct the loss function based on the metrics. 
On one hand, such labels can be costly to obtain (e.g., the high cost of human annotations~\cite{snow2008cheap,chapelle2009expected}) and suffer from biases (e.g., position bias, trust bias, and quality-of-context bias in click logs~\cite{joachims2007evaluating,joachims2017accurately}). 
On the other hand, optimizing ranking metrics based on these fine-grained labels may not always align directly with the ultimate goals of ranking systems, such as user satisfaction and engagement, or other complicated targets like provider-side fairness.

Compared to pointwise fine-grained labels, coarse-grained labels at list level, such as query reformation, search result page re-examination, search duration time, and session engagement duration, are more easily obtainable from search logs and naturally reflect user behaviors and preferences~\cite{dupret2013absence,liu2014skimming,joachims2007evaluating,joachims2017accurately}. 
These signals offer a rich and scalable alternative to expensive human annotations or noisy implicit feedback. 
However, how to incorporate such coarse-grained rewards into the training of LTR methods effectively remains an open question. 
Since these rewards do not provide direct supervision at the item level, they are not compatible with conventional supervised learning paradigms.
This mismatch calls for a different learning paradigm—Reinforcement Learning (RL)—which can directly optimize models using reward signals that reflect overall system performance or user satisfaction, even when fine-grained labels are unavailable. 
Previous studies have explored the use of multiple types of RL algorithms in the LTR task~\cite{wei2017reinforcement,xin2020self,zhou2020rlirank,singh2019policy,oosterhuis2018ranking,chen2019top}.
Unlike supervised learning that relies on explicitly labeled relevance scores to define losses, RL enables training from aggregate feedback through interaction with the environment or from logged behavioral data. 
Thus, RL provides a natural framework to leverage coarse-grained labels as training signals for ranking models.

Despite the promising potential of Reinforcement Learning to Rank (RLTR), most existing approaches mentioned above have primarily employed basic RL algorithms such as vanilla REINFORCE~\cite{sutton1999policy,williams1992simple}.
These algorithms often suffer from high variance in gradient estimation and limited sample efficiency, which can hinder their effectiveness when applied to complex ranking environments with sparse or complex reward signals~\cite{xu2020reinforcement,singh2019policy}.
Meanwhile, recent advances in RL, especially those spurred by the rapid development of large language models (LLMs)~\cite{guo2025deepseek,yang2025qwen3,hurst2024gpt}, have introduced a new generation of RL algorithms with improved stability and performance.
In particular, methods such as Group Relative Policy Optimization (GRPO) and its variants have demonstrated strong empirical success in aligning model outputs with long-horizon, coarse-grained objectives in LLM fine-tuning~\cite{guo2025deepseek,shao2024deepseekmath}. 
Inspired by these advancements, in this paper, we re-examine existing RL methods for LTR and investigate the question: \textit{what's the performance of state-of-the-art RL algorithms in ranking optimization and to what extent RLTR with coarse-grained rewards can replace supervised LTR methods based on fine-grained annotations? } 
Specifically, we explore how to leverage advanced policy optimization algorithms and coarse-grained reward signals to train ranking models that better align with both relevance metrics and system-level goals.

In this work, we systematically explore the performance of traditional fine-grained supervised learning methods and RL algorithms—particularly state-of-the-art RL approaches—under various types of evaluation metrics, model architectures, and coarse-grained reward signals, including classical relevance-based metrics such as NDCG and complex system-level objectives such as ranking fairness. 
Our experiments on three widely used large-scale learning-to-rank datasets, i.e., MSLR 30k, Yahoo! LETOR, and Istella-S LETOR~\cite{lucchese2016post,chapelle2011yahoo,qin2013introducing}, indicate that, even when only coarse-grained reward signals are available, advanced RL algorithms can outperform traditional supervised learning methods on relatively simple relevance-based objectives. 
They can also be used to optimize complex list-level or even system-level ranking objectives such as ranking fairness. 

In summary, this paper makes three key contributions:
(1) We systematically evaluate the effectiveness of existing RLTR methods across diverse ranking architectures and reward types, including relevance metrics like NDCG and fairness objectives to benchmark their performance in ranking optimization with coarse-grained rewards.
(2) We introduce advanced RL algorithms, i.e., GRPO, from LLMs to LTR.
(3) Experimental results demonstrate that RL methods based on coarse-grained signals exhibit strong performance and effectiveness: they not only achieve competitive results on traditional relevance metrics but also show great potential in optimizing more complex global metrics like fairness.
\section{Related Work}
In this section, we first briefly review existing works on RL and its application in LTR, then we discuss the comparison between coarse-grained and fine-grained labels in LTR.

\subsection{Reinforcement Learning to Rank (RLTR)}
Reinforcement Learning (RL) is a well-established research field with applications in various domains such as video games~\cite{shao2019survey}, robotics~\cite{kober2013reinforcement}, and autonomous systems~\cite{arulkumaran2017deep}. 
In particular, RL algorithms have played a pivotal role in enabling capabilities such as alignment and reasoning in the rapidly advancing large language models (LLMs)~\cite{guo2025deepseek,shao2024deepseekmath,yang2025qwen3,hurst2024gpt,achiam2023gpt}.
On the other hand, RL has also gained significant attention in the Information Retrieval (IR) community, particularly in areas like Recommender Systems~\cite{chen2019top,xin2020self,zhao2018deep}, Advertisement~\cite{cai2017real}, and Relevance Feedback~\cite{montazeralghaem2020reinforcement}. 
Some works have also studied RL for LTR tasks, often formulating them as a Markov Decision Process (MDP)~\cite{bellman1957markovian}. 
For instance, \citet{wei2017reinforcement} propose an approach that treats ranking as an MDP and utilizes pointwise learning to build RL models for candidate ranking. 
\citet{xu2020reinforcement} extend this approach by implementing a pairwise policy gradient method to enhance ranking performance further. 
\cite{singh2019policy} and \cite{oosterhuis2021computationally} optimize the relevance and fairness of ranking results using the vanilla policy gradient algorithm along with carefully designed sampling mechanisms. 
However, to the best of our knowledge, there has been limited research on RL with coarse-grained rewards for LTR tasks, specifically in constructing effective ranking models using rewards collected at the rank list level.
Notably, a similar work is \cite{oosterhuis2018ranking} that proposes training an RLTR model with listwise rewards in complex presentation layouts. 
However, it primarily focuses on modeling the ranking setup within complex layouts and on the architecture of the ranking model.
Our work, on the other hand, aims to systematically explore whether RL, particularly recently proposed advanced algorithms, can effectively optimize ranking models with different types of coarse-grained labels.

\subsection{Coarse-grained vs Fine-grained Labels}
In this work, we aim to explore the effectiveness of using coarse-grained labels in training LTR models. 
Most existing LTR works focus on fine-grained labels, which are based on candidate-level relevance judgments. 
However, obtaining such explicit relevance judgments can be costly~\cite{chapelle2009expected,snow2008cheap}. 
To address this issue, some studies have proposed using implicit feedback from search sessions, such as clicks and purchases~\cite{joachims2007evaluating,joachims2017accurately,yadav2020fair}. 
While these implicit feedbacks can provide abundant labels, they suffer from inherent biases, including position bias, trust bias, and quality-of-context bias. 
To mitigate these biases, various unbiased learning to rank (ULTR) algorithms~\cite{joachims2017unbiased,ai2018unbiased,oosterhuis2018differentiable} have been proposed. 
An alternative approach is to bypass these inherent problems associated with fine-grained labels and use coarse-grained labels~\cite{ai2023metric}. 
For example, \citet{oosterhuis2018ranking} attempt to train LTR models with Search Engine Result Page-level (SERP-level) rewards. 
\citet{hu2018reinforcement} propose to use session-level rewards to train RLTR models, where they introduce a session-level MDP to rank items based on user behaviors within a session dynamically. 
In this work, since we do not have access to real-world search logs, we are unable to compare our model with theirs directly. 
Instead, we adopt a similar approach as~\cite{oosterhuis2018ranking}, which involves simulating listwise labels from candidate-level relevance judgments and utilizing these simulated coarse-grained labels to train our LTR models. 
It is worth noting that our work can be seamlessly extended to utilize other coarse-grained rewards or metrics discussed earlier.
\section{Preliminaries}
In this section, we first formalize the problem by framing coarse-grained ranking as a listwise learning task, where the goal is to generate the entire ranking list in a single step. 
We further introduce the modeling approach based on the Plackett-Luce model.

\begin{figure}[t]
    \centering
    \includegraphics[width=0.48\textwidth]{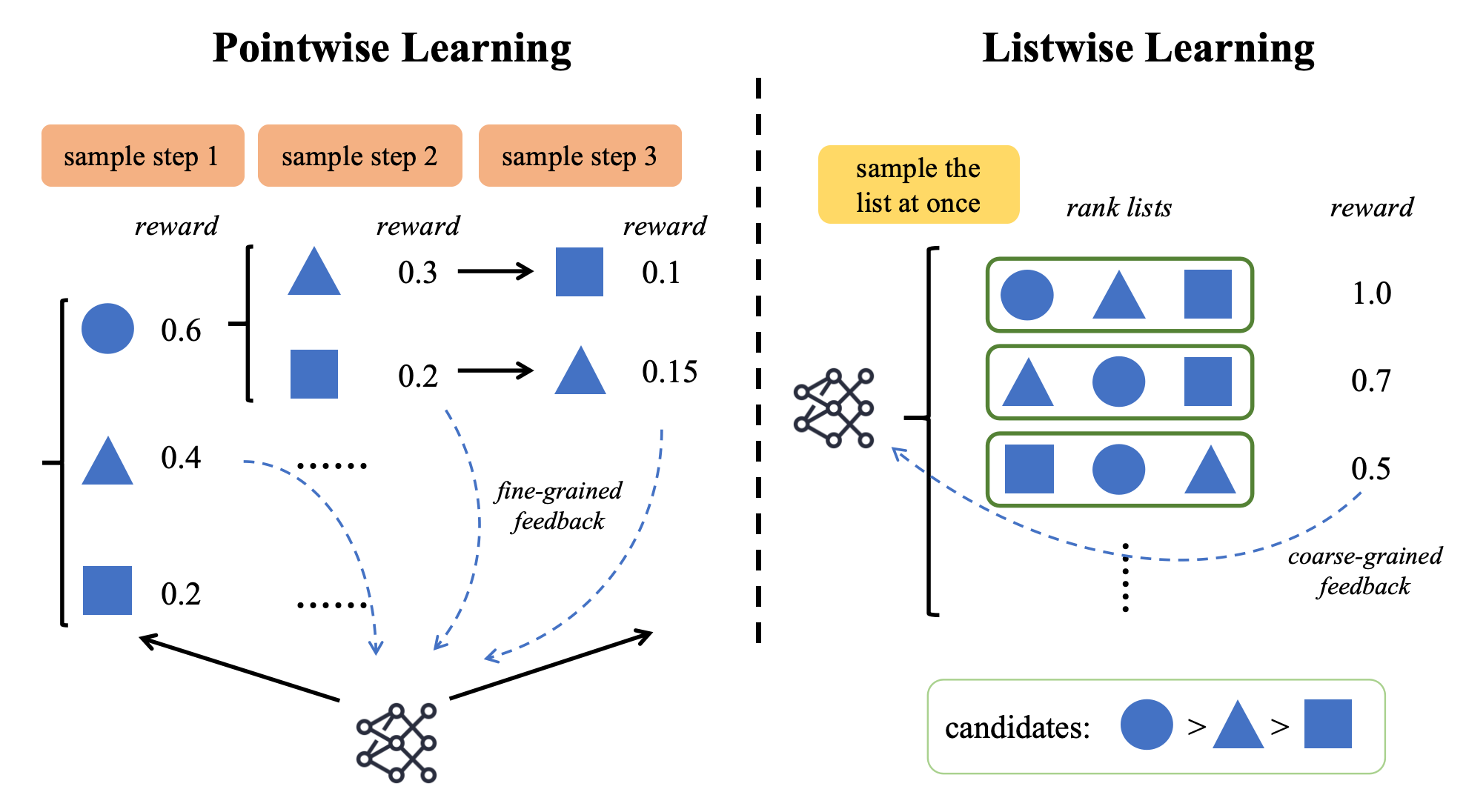}
    \caption{An illustration of pointwise learning with fine-grained labels and listwise learning with coarse-grained rewards in RLTR.}
    \label{fig:point_and_list}
\end{figure}

\subsection{Problem Formulation}
We aim to investigate whether we can train a ranking model with coarse-grained listwise rewards $\mathcal{R}$ to effectively rank candidates (e.g., query-document pairs in search engines, user-item pairs in recommendation systems, etc.) based on their input feature vectors  $X = (x_1, x_2, ..., x_k)$, where each $x_i$ represents the feature vector of one ranking candidate, and $k$ denotes the size of the rank list.
The ultimate goal of the model is to learn a ranking policy $\pi$ that maximizes the reward $\mathcal{R}(L)$ from the rank list $L$.
Generally, RLTR approaches can be divided into two learning methods: pointwise learning and listwise learning (as shown in Figure~\ref{fig:point_and_list}). 
In pointwise learning, ranking is regarded as a sequence of actions where the model selects the appropriate document or item for each slot in the rank list~\cite{wei2017reinforcement,xu2020deep}. 
Thus, a rank list containing $k$ ranking candidates results from $k$ discrete ranking time steps. 
Although this approach takes advantage of the discrete nature of classical RL (i.e., only picking one action per time step), it always requires fine-grained candidate-level rewards. 
However, when such rewards are unavailable, we have to explore alternative methods. 

On the contrary, listwise learning directly generates a rank list corresponding to the query~\cite{singh2019policy,oosterhuis2018ranking}. 
It constructs a loss function based on listwise rewards collected from the rank list, meaning that the ranking model can be directly trained via coarse-grained listwise rewards. 
Besides, generating the complete list for optimization directly can help avoid limitations imposed by the structure or characteristics of coarse-grained reward functions~\cite{singh2019policy}, which offers superior generalization capability. 
Therefore, we model the ranking process by generating the entire rank list $L$ at once, rather than selecting candidates sequentially at each time step.

\subsection{Plackett-Luce Ranking Model}
The Plackett-Luce (PL) ranking model ~\cite{plackett1975analysis,luce1959individual} is a classical probabilistic ranking model that defines a distribution over permutations based on item scores.
It is grounded in the principle of sequential choice, assuming that a ranking is formed by repeatedly selecting the next item according to a multinomial distribution over the remaining candidates.
Thus, it defines a distribution over the entire space of ranking permutations, making it suitable for our settings where listwise supervision is available.
Formally, given a scoring model $h_\theta$ with learnable parameters $\theta$, a list of input vectors $X$, the output of the model is a score vector of size $k$:
\begin{equation}
    h_\theta(X) = \left(h_\theta(x_1),h_\theta(x_2),...,h_\theta(x_{k})\right).
\end{equation}

Based on this vector, we can derive the probability $\pi_\theta(L \thinspace | \thinspace s)$ of generating a sampled rank list $L=(L[1],...,L[k])$ as follow:
\begin{equation}
\label{eq:plackett}
\pi_{\theta}(L \mid X)=\prod_{i=1}^{k} \frac{\exp \left(h_{\theta}(x_{L[i]})\right)}{\exp \left(h_{\theta}(x_{L[i]})\right)+\ldots+\exp \left(h_{\theta}(x_{L[k]})\right)}.
\end{equation}

While the PL model provides a principled probabilistic framework for modeling rankings, exact sampling from the PL distribution, especially when repeated sampling is required, can be computationally expensive due to its sequential nature.
To address this issue, we can apply the Gumbel Softmax trick~\cite{gumbel1954statistical} to enable efficient and parallelizable sampling from the PL distribution with a computational complexity
of $\mathcal{O}\left(k\log k\right)$~\cite{oosterhuis2021computationally,bruch2020stochastic}. 
Instead of directly calculating the placement probabilities, it adds independent Gumbel noise to the original ranking scores:
\begin{equation}
    \hat{h}_\theta(x_i) = h_\theta(x_i) + \gamma_i,
\end{equation}
where $\gamma_i \sim \text{Gumbel}(0, 0)$ is the Gumbel noise, and a sampled ranking $L$ is obtained by sorting the candidates in descending order of their perturbed scores $\hat{h}_\theta(x_i)$.

\section{Methods}
In this section, we first introduce three ranking methods based on classical reinforcement learning algorithms: PGRank, PPG, and PLRank, which represent the classic RLTR algorithms in the literature.
We then explore the application of a modern and advanced RL algorithm, Group Relative Policy Optimization (GRPO), to coarse-grained ranking tasks.

\subsection{Classic RLTR Algorithms}
\subsubsection{PGRank~\cite{singh2019policy}}
PGRank formulates LTR as a policy learning problem, where the ranking function is treated as a stochastic policy optimized via policy gradient methods. 
By modeling the ranking process through a PL distribution and employing REINFORCE-based~\cite{sutton1999policy,williams1992simple} updates, PGRank directly maximizes any desired user-utility metric  like NDCG:
\begin{equation}\label{eq:pgrank}
J_{\text{PGRank}}(\theta) = \mathbb{E}_{L \sim \pi_{\theta}}
\big[\mathcal{R}(L) \cdot \nabla_{\theta} \log \underbrace{\pi_{\theta}(L \mid X)}_{\mathrm{Eq}.~(\ref{eq:plackett})}\big].
\end{equation}

Note that to reduce the variance of the algorithm, the listwise reward function $R(L)$ is adjusted by subtracting a baseline term $b(q)$, which is computed as the average reward over multiple sampled lists for the same query $q$.

\subsubsection{PPG~\cite{xu2020reinforcement}}
To fully utilize the relative relationships between candidates within the same ranking session and alleviate the problem of high variance gradient estimation, PPG proposes an intra-query pairwise sampling strategy.
Two rank lists are sampled from the same state, and their performance difference is used to construct the gradient.
Note that the original PPG model aims at optimizing the process of ranking items one by one, but in our setting, we need to generate the entire rank list at once. Therefore, we still use the PL distribution to model the policy:
\begin{equation}\begin{aligned}
J_{\text{PPG}}(\theta) = \mathbb{E}_{L_1, L2 \sim \pi_\theta} \bigg[\big( \mathcal{R}(L_1) - \mathcal{R}(L_2)\big) \cdot & \\  \big( \nabla_\theta \log \pi_\theta(L_1\ |\ X) - & \nabla_\theta \log \pi_\theta(L_2\ |\ X) \big) \bigg],
\end{aligned}\end{equation}
where $L_1, L_2$ denote two rank lists sampled on the same query, while the policy $\pi_\theta$ is also computed by $\mathrm{Eq}.~(\ref{eq:plackett})$.

\subsubsection{PLRank~\cite{oosterhuis2021computationally}} 
PLRank estimates the gradient of a PL ranking model w.r.t. relevance or fairness metrics at the fine-grained level.
Specifically, it makes use of the specific structure of the PL model and the objective function and decomposes the sampling strategy and reward function at the fine-grained level, enabling RL based on candidate-level rewards.
Therefore, it is a pointwise RL algorithm trained with \textbf{fine-grained} reward signals. 
Specifically, its loss function can be simplified as follows (note that we omit some complicated acceleration tricks in the original paper):
\begin{equation}\begin{aligned}
J_{\text{PLRank}}(\theta) = \mathbb{E}_{L \sim \pi_{\theta}}
\bigg[\sum_{i=1}^k \nabla_{\theta} \log \pi_{\theta}\big(L[i] \mid X, L[1:i-1]\big) \\ \cdot \sum_{j=i}^k \mathcal{R}(L[j])\bigg],
\end{aligned}\end{equation}
where $\pi_{\theta}\big(L[i] \mid X, L[1:i-1]\big)$ denotes the probability of selecting the $i$-th item given the first $i-1$ items in the list, and $\mathcal{R}(L[j])$ is the candidate-level reward.

\subsection{Group Relative Policy Optimization (GRPO)}
Recently, RL has emerged as a critical component in training large language models, particularly for enhancing reasoning capabilities~\cite{guo2025deepseek,hurst2024gpt,yang2025qwen3}.
Group Relative Policy Optimization (GRPO)~\cite{guo2025deepseek,shao2024deepseekmath} is among the most representative algorithms in this context.
As a PPO~\cite{schulman2017proximal}-inspired, critic‑free approach, GRPO samples multiple outputs from the same input, computes a group-normalized reward advantage $\hat{A}$, and updates the policy using a clipped surrogate objective with KL regularization—altogether avoiding the need for a value function and reducing memory and computational overhead.
The optimization objective of the original GRPO algorithm is: 
\begin{equation}\begin{aligned}\label{eq:grpo_original}
J_{\text{GRPO}}(\theta) &= \mathbb{E}_{L_i \sim \pi_{\theta_{old}}} \frac{1}{G}\sum_{i=1}^G \bigg[\min \big\{r_\theta(X, L_i) \cdot \hat{A}(L_i), \\ &\text{clip}\big(r_\theta(X, L_i), 1 - \epsilon, 1 + \epsilon\big) \cdot \hat{A}(L_i)\big\} - \beta \mathbb{D}_{KL}(\pi_\theta || \pi_{ref})\bigg],
\end{aligned}\end{equation}
where the group size $G$ denotes the number of permutations sampled under the same query, $r_\theta(X, L_i) = \frac{\pi_{\theta}(L_i\ \mid X)}{\pi_{\theta_{old}}(L_i\ \mid X)}$ is the importance ratio between the current policy $\pi_\theta$ and old policy $\pi_{\theta_{old}}$, and the advantage function $\hat{A}(L_i)$ is computed based on the group-normalized reward:
\begin{equation}
\hat{A}(L_i) = \frac{\mathcal{R}(L_i) - \text{mean}\left(\{\mathcal{R}(L_i)\}_{i=1}^G\right)}{\text{std}\left(\{\mathcal{R}(L_i)\}_{i=1}^G\right) + \delta},
\end{equation}
where $L_i$ denotes the rank list sampled under the same query $q$, and $\delta$ is a small constant to avoid division by zero. The KL divergence $\mathbb{D}_{KL}(\pi_\theta || \pi_{ref})$ penalizes the deviation between the new policy and the frozen reference policy.
Unlike the training LLMs setting, where the model already possesses certain language and reasoning capabilities, the ranking score model in our case is trained from scratch. 
Therefore, instead of using a pre-trained model as the reference, we define the reference policy as a checkpoint saved every fixed number of training steps.
The clipping operation $\text{clip}(\cdot)$ is used to constrain the update ratio of the new policy relative to the old policy, which helps improve training stability to some extent. 
However, this also requires the model to be updated multiple times on the same batch of data sampled from $\pi_{\theta_{old}}$, which limits training efficiency. 
Therefore, in this work, we only consider a one-step update strategy, under which the optimization objective can be simplified as:
\begin{equation}\begin{aligned}
J_{\text{GRPO}}(\theta) = \mathbb{E}_{L_i \sim \pi_{\theta}} \frac{1}{G}\sum_{i=1}^G \big[ & \frac{\pi_{\theta}(L_i\ \mid X)}{[\pi_{\theta}(L_i\ \mid X)]_{\text{no grad}}}  \cdot \hat{A}(L_i) \\ & \qquad \qquad - \beta \mathbb{D}_{KL}(\pi_\theta || \pi_{ref})\big].
\end{aligned}\end{equation}

\section{Experimental Settings}
\subsection{Datasets \& Evaluation Metrics}

\begin{table}[t]
    \centering
    \caption{A summary of the dataset statistics.}
    \resizebox{0.95\columnwidth}{!}{
    \begin{tabular}{c|ccc}
    \toprule
    Property/Dataset & MSLR 30k & Yahoo! & Istella-S \\
    \midrule
    \#Queries & 31,151 & 29,921 & 33,018 \\
    \#Documents & 376k & 71k & 340k \\
    \#Avg. documents per query & 121 & 24 & 103 \\
    \#Features & 136 & 700 & 220 \\
    \#Label level & 5 & 5 & 5 \\  
    \bottomrule
    \end{tabular}
    }
    \label{tab:Dataset_statistic}
\end{table}

We conduct experiments on three standard LTR datasets: Microsoft Learning to Rank Datasets Fold 1 (MSLR 30k, MSLR)\footnote{\url{https://www.microsoft.com/en-us/research/project/mslr/}}, Yahoo! LETOR Challenge set 1 (Yahoo!)\footnote{\url{https://webscope.sandbox.yahoo.com}} and Istella-S LETOR (Istella-S)\footnote{\url{http://blog.istella.it/istella-learning-to-rank-dataset/}}~\cite{qin2013introducing,chapelle2011yahoo,lucchese2016post}. 
All datasets include multiple candidate lists (each represented with a query ID) and the corresponding features for each candidate (i.e., query-document pair). 
The specific properties of the three datasets are outlined in Table~\ref{tab:Dataset_statistic}. 
We follow the data preprocessing pipeline from~\cite{tran2021ultra}.

We primarily evaluate our approaches from two perspectives: relevance and fairness.
We select two standard metrics: Normalized Discounted Cumulative Gain (NDCG)~\cite{jarvelin2002cumulated} and Expected Reciprocal Rank (ERR) ~\cite{chapelle2009expected}, to measure the performance of each algorithm in relevance experiments. 
NDCG is a metric based on information gain theory, while ERR is a metric capturing users' satisfaction with search results. 
During training, we select the checkpoint that performs the best on the validation set based on the metric score. 
We report the average scores of these metrics for the top-3 and top-10 candidates. 

Regarding the fairness experiment, we directly adopt one of the most popular fairness metrics, i.e., individual exposure fairness, as our metric.  
Specifically, it considers the exposure disparity between candidate pairs in a rank list as the individual exposure unfairness measure~\cite{oosterhuis2021computationally, yang2021maximizing}:
\begin{equation}\label{eq:fair}
fair(q)=-unfair(q),
\end{equation}
\begin{equation}\label{eq:unfair}
unfair(q) = \frac{1}{k(k - 1)} \sum_{d_x, d_y \in L_q} \left(E(d_x)R(d_y) - E(d_y)R(d_x)\right)^2,
\end{equation}
\begin{equation}\label{eq:exposure}
E(d) = \sum_{\tau = 1}^T\sum_{i=1}^k\frac{1}{\log_2(i + 1)}\mathbb{I}\big(L_{q, \tau}[i] == d\big),
\end{equation} 
\begin{equation}
R(d) = \frac{2^{y_d} - 1}{2^{y_{\rm{max}}} - 1},
\end{equation} 
where $E(\cdot)$ represents the exposure function, $T=100$ denotes the number of samples for exposure calculation, and $L_{q, \tau}$ refers to the rank list at the $\tau$-th sampling w.r.t. query $q$. 
When it comes to the relevance function $R(\cdot)$, $y_d$ represents the relevance judgment of $d$, while $y_{\rm{max}}$ represents the maximum relevance judgment of the entire dataset.
Note that in fairness experiments, since the number of candidates for different queries varies, we ensure consistent dimensions by keeping only the top 10 candidates from the initial rank (given by SVM-rank following~\cite{tran2021ultra}) for each query.

\subsection{Simulated Coarse-grained Rewards} 
To simulate the coarse-grained rewards in the relevance experiment, we primarily use the provided ground truth labels to calculate the list-level relevance metric on the top-10 results (e.g., NDCG@10 and ERR@10) on the rank list generated by the models. 
This allows us to simulate the feedback received from online sessions.

For the fairness reward function, we adopt the exposure-gradient-based signal from~\citet{oosterhuis2021computationally}. Specifically, a listwise fairness reward is calculated as:
\begin{equation}
\mathcal{R}^{fair}(L) = \sum_{i=1}^k \frac{\rho^{fair}_{L[i]}}{\log_2(i + 1)},
\end{equation}
where $\rho$ denotes the fairness utility of a candidate. It is defined as the partial derivative of fairness~(Eq.~(\ref{eq:fair}), Eq.~(\ref{eq:unfair})) w.r.t. the exposure function~(Eq.~(\ref{eq:exposure})):
\begin{equation}\begin{aligned}
\rho^{fair}_d & = \frac{\partial \ fair(q)}{\partial \ E(d)} \\& = \frac{4}{k(k-1)}  \sum_{d' \in L_q} \left(E(d')R(d) - E(d)R(d')\right)R(d').
\end{aligned}\end{equation}

\subsection{Baselines}

\begin{table*}[htbp]
\renewcommand{\arraystretch}{1.2}
\centering
\caption{The performance of baseline supervised learning methods and RL approaches that use NDCG@10 as the reward function in relevance experiments, across different datasets and backbone ranking models. The best and second-best methods are marked in bold and underlined, respectively. "$\dagger$" and "$\ddagger$" indicate significantly worse than the best and second-best method at the $p < 0.05$ level using the two-tailed pairwise t-test, respectively.}
\label{tab:ndcg_result}
\scalebox{0.75}{
\begin{tabular}{c|c|cccc|cccc|cccc}
\toprule
 & Model & \multicolumn{4}{c|}{MLP} & \multicolumn{4}{c|}{DLCM} & \multicolumn{4}{c}{SetRank} \\
 \cline{2-14}
\multirow{-2}{*}{Dataset} & Algorithm / Metrics & ERR@3 & NDCG@3 & ERR@10 & NDCG@10 & ERR@3 & NDCG@3 & ERR@10 & NDCG@10 & ERR@3 & NDCG@3 & ERR@10 & NDCG@10 \\ 
\midrule
 & CrossEntropy & 0.3009$^{\dagger\ddagger}$ & 0.4238$^{\dagger\ddagger}$ & 0.3436$^{\dagger\ddagger}$ & 0.4465$^{\dagger\ddagger}$ & 0.2588$^{\dagger\ddagger}$  & 0.3737$^{\dagger\ddagger}$  & 0.3045$^{\dagger\ddagger}$  & 0.4092$^{\dagger\ddagger}$  & 0.2693$^{\dagger\ddagger}$ & 0.3860$^{\dagger\ddagger}$ & 0.3135$^{\dagger\ddagger}$ & 0.4149$^{\dagger\ddagger}$ \\
 & AttentionRank & \textbf{0.3120} & \underline{0.4274} & \textbf{0.3544} & \underline{0.4500} & \textbf{0.2774} & \textbf{0.3882} & \textbf{0.3212} & \textbf{0.4185} & \underline{0.2808}$^{\dagger}$ & \underline{0.3919}$^{\dagger}$ & \underline{0.3234}$^{\dagger}$ & \underline{0.4177}$^{\dagger}$ \\
 & LambdaRank & 0.3084$^{\dagger\ddagger}$ & 0.4255$^{\dagger}$ & 0.3508$^{\dagger\ddagger}$ & 0.4481$^{\dagger\ddagger}$ & 0.2619$^{\dagger\ddagger}$  & 0.3770$^{\dagger\ddagger}$  & 0.3072$^{\dagger\ddagger}$  & 0.4098$^{\dagger\ddagger}$  & 0.2757$^{\dagger\ddagger}$ & 0.3884$^{\dagger\ddagger}$ & 0.3189$^{\dagger\ddagger}$ & 0.4157$^{\dagger\ddagger}$ \\
  & PLRank & $0.2723^{\dagger\ddagger}$ & $0.3832^{\dagger\ddagger}$ & $0.3159^{\dagger\ddagger}$ & $0.4128^{\dagger\ddagger}$ & $0.2374^{\dagger\ddagger}$ & $0.3428^{\dagger\ddagger}$ & $0.2831^{\dagger\ddagger}$ & $0.3783^{\dagger\ddagger}$ & $0.2559^{\dagger\ddagger}$ & $0.3631^{\dagger\ddagger}$ & $0.3003^{\dagger\ddagger}$ & $0.3958^{\dagger\ddagger}$ \\
 \cline{2-14}
 & PGRank & 0.3042$^{\dagger\ddagger}$ & 0.4178$^{\dagger\ddagger}$ & 0.3470$^{\dagger\ddagger}$ & 0.4415$^{\dagger\ddagger}$ & 0.2561$^{\dagger\ddagger}$  & 0.3679$^{\dagger\ddagger}$  & 0.3016$^{\dagger\ddagger}$  & 0.4020$^{\dagger\ddagger}$  & 0.2660$^{\dagger\ddagger}$ & 0.3790$^{\dagger\ddagger}$ & 0.3101$^{\dagger\ddagger}$ & 0.4094$^{\dagger\ddagger}$ \\
 & PPG & 0.3011$^{\dagger\ddagger}$ & 0.4154$^{\dagger\ddagger}$ & 0.3437$^{\dagger\ddagger}$ & 0.4388$^{\dagger\ddagger}$ & 0.2555$^{\dagger\ddagger}$  & 0.3677$^{\dagger\ddagger}$  & 0.3009$^{\dagger\ddagger}$  & 0.4012$^{\dagger\ddagger}$  & 0.2651$^{\dagger\ddagger}$ & 0.3766$^{\dagger\ddagger}$ & 0.3095$^{\dagger\ddagger}$ & 0.4076$^{\dagger\ddagger}$ \\
\multirow{-7}{*}{MSLR 30k} 
 & GRPO & \underline{0.3104} & \textbf{0.4295} & \underline{0.3528} & \textbf{0.4510} & \underline{0.2739} & \underline{0.3869} & \underline{0.3182} & \underline{0.4181} & \textbf{0.2947} & \textbf{0.4079} & \textbf{0.3375} & \textbf{0.4324} \\

\midrule
 & CrossEntropy & 0.6956$^{\dagger\ddagger}$ & 0.6235$^{\dagger\ddagger}$  & 0.7190$^{\dagger\ddagger}$ & \underline{0.7114} & 0.6693$^{\dagger\ddagger}$ & 0.5914$^{\dagger\ddagger}$ & 0.6957$^{\dagger\ddagger}$ & 0.6800$^{\dagger}$ & $0.6853^{\dagger}$ & $0.6121^{\dagger}$ & $0.7099^{\dagger}$ & $\underline{0.6966}^{\dagger}$ \\
 & AttentionRank & \underline{0.7017} & \underline{0.6301} & \underline{0.7250} & 0.7113 & \underline{0.6754} & \underline{0.5973}$^{\dagger}$ & \underline{0.7012} & \underline{0.6819}$^{\dagger}$ & $0.6851^{\dagger\ddagger}$ & $0.6121^{\dagger}$ & $0.7098^{\dagger\ddagger}$ & $0.6963^{\dagger}$ \\
 & LambdaRank & 0.7000 & 0.6286$^{\dagger}$  & 0.7229$^{\dagger}$ & 0.7078$^{\dagger\ddagger}$ & 0.6682$^{\dagger\ddagger}$ & 0.5894$^{\dagger\ddagger}$ & 0.6939$^{\dagger\ddagger}$ & 0.6735$^{\dagger\ddagger}$ & $\underline{0.6894}^{\dagger}$ & $\underline{0.6131}^{\dagger}$ & $\underline{0.7134}^{\dagger}$ & $0.6900^{\dagger\ddagger}$ \\
  & PLRank & $0.6782^{\dagger\ddagger}$ & $0.5997^{\dagger\ddagger}$ & $0.7032^{\dagger\ddagger}$ & $0.6756^{\dagger\ddagger}$ & $0.6309^{\dagger\ddagger}$ & $0.5457^{\dagger\ddagger}$ & $0.6596^{\dagger\ddagger}$ & $0.6221^{\dagger\ddagger}$ & $0.6666^{\dagger\ddagger}$ & $0.5844^{\dagger\ddagger}$ & $0.6919^{\dagger\ddagger}$ & $0.6572^{\dagger\ddagger}$ \\
 \cline{2-14}
 & PGRank & 0.6962$^{\dagger\ddagger}$ & 0.6224$^{\dagger\ddagger}$ & 0.7198$^{\dagger\ddagger}$ & 0.7001$^{\dagger\ddagger}$ & 0.6638$^{\dagger\ddagger}$ & 0.5859$^{\dagger\ddagger}$ & 0.6939$^{\dagger\ddagger}$ & 0.6735$^{\dagger\ddagger}$ & $0.6716^{\dagger\ddagger}$ & $0.5991^{\dagger\ddagger}$ & $0.6963^{\dagger\ddagger}$ & $0.6735^{\dagger\ddagger}$ \\
 & PPG & 0.6936$^{\dagger\ddagger}$ & 0.6192$^{\dagger\ddagger}$ & 0.7179$^{\dagger\ddagger}$ & 0.6984$^{\dagger\ddagger}$ & 0.6563$^{\dagger\ddagger}$ & 0.5781$^{\dagger\ddagger}$ & 0.6832$^{\dagger\ddagger}$ & 0.6604$^{\dagger\ddagger}$ & $0.6793^{\dagger\ddagger}$ & $0.6028^{\dagger\ddagger}$ & $0.7039^{\dagger\ddagger}$ & $0.6793^{\dagger\ddagger}$ \\
\multirow{-7}{*}{Istella-S} 
 & GRPO & \textbf{0.7030} & \textbf{0.6329} & \textbf{0.7255} & \textbf{0.7124} & \textbf{0.6768} & \textbf{0.6009} & \textbf{0.7025} & \textbf{0.6873} & \textbf{0.6938} & \textbf{0.6197} & \textbf{0.7176} & \textbf{0.7004}  \\

\midrule
 & CrossEntropy & $0.4227^{\dagger\ddagger}$ & $0.6860^{\dagger\ddagger}$ & $0.4601^{\dagger\ddagger}$ & $0.7558^{\dagger\ddagger}$ & $0.4124^{\dagger\ddagger}$ & $0.6707^{\dagger\ddagger}$ & $0.4504^{\dagger\ddagger}$ & $0.7452^{\dagger\ddagger}$ & $0.4222^{\dagger\ddagger}$ & $0.6824^{\dagger\ddagger}$ & $0.4598^{\dagger\ddagger}$ & $0.7538^{\dagger\ddagger}$ \\
 & AttentionRank & $0.4291^{\dagger}$ & $0.6924^{\dagger}$ & $0.4659^{\dagger}$ & $0.7613^{\dagger}$ & $\underline{0.4190}^{\dagger}$ & $\underline{0.6789}^{\dagger}$ & $\underline{0.4565}^{\dagger}$ & $\underline{0.7509}^{\dagger}$ & $0.4280^{\dagger}$ & $0.6911^{\dagger\ddagger}$ & $0.4649^{\dagger}$ & $0.7581^{\dagger\ddagger}$ \\
 & LambdaRank & $\underline{0.4298}^{\dagger}$ & $\underline{0.6961}^{\dagger}$ & $0.4663^{\dagger}$ & $0.7618^{\dagger}$ & $0.4151^{\dagger\ddagger}$ & $0.6762^{\dagger}$ & $0.4530^{\dagger\ddagger}$ & $0.7487^{\ddagger}$ & $\underline{0.4289}^{\dagger}$ & $\underline{0.6950}^{\dagger}$ & $\underline{0.4656}^{\dagger}$ & $\underline{0.7603}^{\dagger}$ \\
  & PLRank & $0.4245^{\dagger\ddagger}$ & $0.6839^{\dagger\ddagger}$ & $0.4617^{\dagger\ddagger}$ & $0.7526^{\dagger\ddagger}$ & $0.4137^{\dagger\ddagger}$ & $0.6639^{\dagger\ddagger}$ & $0.4515^{\dagger\ddagger}$ & $0.7361^{\dagger\ddagger}$ & $0.4150^{\dagger\ddagger}$ & $0.6635^{\dagger\ddagger}$ & $0.4531^{\dagger\ddagger}$ & $0.7381^{\dagger\ddagger}$ \\
 \cline{2-14}
 & PGRank & $0.4292^{\dagger}$ & $0.6940^{\dagger}$ & $0.4661^{\dagger}$ & $0.7617^{\dagger}$ & $0.4166^{\dagger}$ & $0.6748^{\dagger}$ & $0.4539^{\dagger\ddagger}$ & $0.7468^{\dagger\ddagger}$ & $0.4282^{\dagger}$ & $0.6912^{\dagger\ddagger}$ & $0.4649^{\dagger}$ & $0.7549^{\dagger\ddagger}$ \\
 & PPG & $0.4297^{\dagger}$ & $0.6955^{\dagger}$ & $\underline{0.4666}^{\dagger}$ & $\underline{0.7621}^{\dagger}$ & $0.4178^{\dagger}$ & $0.6748^{\dagger}$ & $0.4552^{\dagger}$ & $0.7469^{\dagger\ddagger}$ & $0.4282^{\dagger}$ & $0.6885^{\dagger\ddagger}$ & $0.4652^{\dagger}$ & $0.7562^{\dagger\ddagger}$ \\
\multirow{-7}{*}{Yahoo!} 
 & GRPO & \textbf{0.4329} & \textbf{0.7023} & \textbf{0.4691} & \textbf{0.7668} & \textbf{0.4260} & \textbf{0.6940} & \textbf{0.4628} & \textbf{0.7618} & \textbf{0.4323} & \textbf{0.7010} & \textbf{0.4686} & \textbf{0.7664} \\
\bottomrule
\end{tabular}
}
\end{table*}

We aim to explore whether ranking models trained with coarse-grained listwise rewards and RL algorithms can achieve comparable or even superior performance to a model trained with fine-grained labels. 
When optimizing ranking relevance, we employ CrossEntropy~\cite{ai2021unbiased,burges2006learning}, AttentionRank~\cite{ai2021unbiased,ai2018learning}, and LambdaRank~\cite{wang2018lambdaloss} as our baseline supervised learning algorithms. 
CrossEntropy and AttentionRank compute attention distributions based on the scores assigned by the ranking model and compare them with attention distributions derived from ground-truth labels.
Both methods are optimized using the cross-entropy loss; the only difference lies in how the attention distribution is computed. 
CrossEntropy directly uses the sum of raw relevance labels as the normalization term in the denominator, whereas AttentionRank applies a softmax operation over the relevance labels to obtain the attention weights.
On the other hand, the loss function of LambdaRank is formulated as:
\begin{equation}\begin{aligned}
\mathcal{J}_{\text{LambdaRank}} & = \sum_{i=1}^{k} \sum_{j: y_{j}<y_{i}} \Delta \text{Rel}(i, j) \cdot \\ & \log_2  \left(1+\exp(-\sigma(h_{\theta}(x_i)-h_{\theta}(x_j)))\right),
\end{aligned}\end{equation}
where $\Delta \text{Rel}(i, j)$ is the absolute difference between the relevance metric (e.g., NDCG or ERR) when two candidates $i, j$ are flipped. 

In fairness experiments, since both PGRank~\cite{singh2019policy} and PLRank~\cite{oosterhuis2021computationally} themselves are effective algorithms for fairness-aware ranking, and the fairness metric is generally difficult to optimize directly via supervised learning methods, we do not introduce additional baseline models.

To demonstrate the effectiveness and generalization ability of our method, we conduct experiments on three distinct neural ranking architectures with varying inductive biases and complexities: a simple Multi-Layer Perceptron (MLP) that models item-wise scoring independently, the Deep Listwise Context Model (DLCM)~\cite{ai2018learning} that incorporates top-down attention over ranked contexts using the GRU module~\cite{cho2014properties}, and SetRank~\cite{pang2020setrank}, a recent permutation-invariant model designed to capture item interactions based on the Transformer architecture and self-attention mechanism~\cite{vaswani2017attention}. 
This diversity of backbone architectures allows us to evaluate whether our method consistently improves performance across different types of ranking models.

\subsection{Training Settings}
To create a more authentic and realistic search engine scenario and make a fair comparison among different models, we simulate an environment where users interact with the search engine, and models are dynamically trained and tested on the fly.
Users can only see the top $k=10$ candidates (sorted by the model) in the list of a given query and provide a feedback signal for the list. 
The feedback rewards are calculated at the coarse-grained and listwise level, simulated with candidate-level annotations.

In relevance experiments, all models are trained with a batch size of 256, an AdamW optimizer~\cite{loshchilov2017decoupled}, and a learning rate of 1e-4 for 10,000 steps. 
For RL algorithms, we sample $G = 8$ rank lists for each query in a training step. 
In the fairness experiments, due to the increased complexity of the reward function, we set a higher learning rate of 1e-2. 
Additionally, we extend the training steps to 100,000 and sample 100 candidate rank lists to promote convergence.
For GRPO, the weight of the KL divergence term is selected from the range $[0, 0.1]$, and during its training, a checkpoint is saved every 500 steps, and the corresponding model is used as the reference model $\pi_{ref}$.
All experiments are conducted on a single \textit{NVIDIA GeForce RTX 3090} GPU.
\section{Results and Analysis}
In this section, we demonstrate our experimental results and summarize our initial investigation of coarse-grained RLTR.
Specifically, we aim to explore the following four research questions thoroughly:
\begin{itemize}[leftmargin=*]
\item \textbf{RQ1:} How do RL algorithms perform on coarse-grained relevance signals (e.g., NDCG) compared to traditional fine-grained supervised learning methods?
\item \textbf{RQ2:} How do RL algorithms perform on complex signals like ranking fairness?
\item \textbf{RQ3:} How robust are the RL algorithms in the presence of noise?
\item \textbf{RQ4:} How does the number of samples in RL affect the performance of RL algorithms?
\item \textbf{RQ5:} Can RL algorithms optimize multiple reward signals simultaneously?
\end{itemize}

\subsection{Relevance Results (RQ1)}
\begin{figure}[t]
    \centering
    \includegraphics[width=0.48\textwidth]{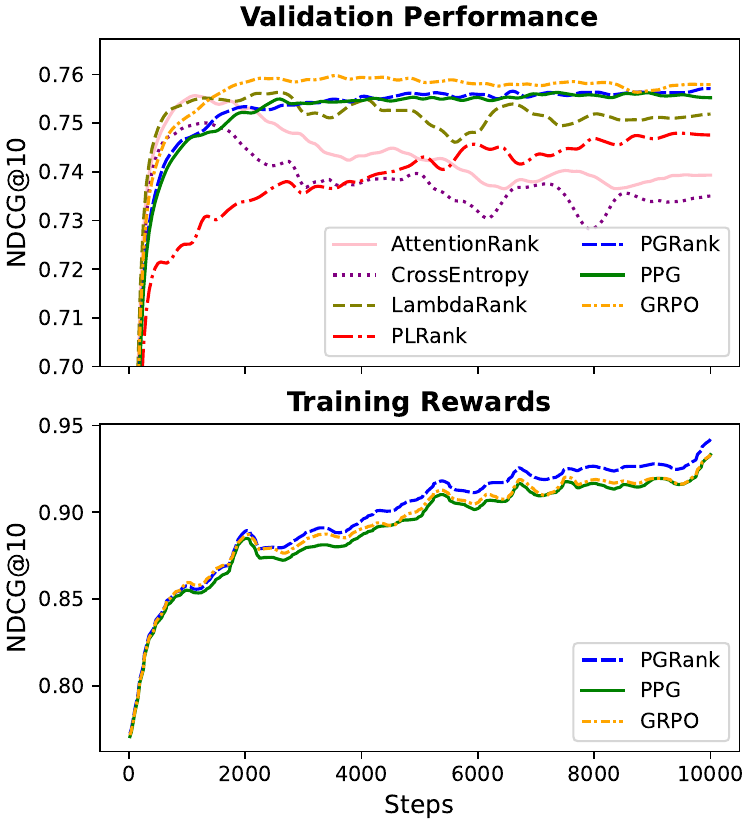}
    \caption{Validation curves of all algorithms (above) and reward trajectories of coarse-grained RL methods (below) on the Yahoo! dataset using MLP as the ranking model.}
    \label{fig:ndcg_curve}
\end{figure}

Table~\ref{tab:ndcg_result} reports all the experimental results when using NDCG@10 as the coarse-grained reward function on the three datasets: MSLR 30k, Yahoo!, and Istella-S, as well as three backbone ranking models, MLP, DLCM, and SetRank.
Despite relying solely on coarse-grained rewards, we surprisingly observe that GRPO consistently outperforms both supervised baselines and other coarse-grained RL methods.
Even the relatively simple coarse-grained RL methods, i.e., PGRank and PPG, while not outperforming the strongest supervised learning approaches, still achieve performance comparable to that of weaker supervised baselines.
Specifically, on Istella-S and Yahoo!, GRPO obtains the best performance in all metrics and across all backbone architectures.
For example, on the Yahoo! dataset and under the DLCM model, GRPO achieves scores of 0.6940 on NDCG@3 and 0.7618 on NDCG@10, outperforming the second-best model by 2.2\% and 1.5\%, respectively.
In MSLR 30k, GRPO still achieves the best or second-best results in all metrics with all backbone models. 
GRPO’s consistent superiority across different architectures and datasets suggests its robustness and generalization capability.
This indicates that, with the aid of powerful RL techniques, training ranking models using coarse-grained reward signals can achieve—and in most cases even surpass—the performance of methods optimized with fine-grained supervision.
On the other hand, the experimental results also reveal the feasibility and strong potential of optimizing ranking models using coarse-grained rewards. 
Even when employing relatively simple RL algorithms, the performance is not necessarily inferior to that of methods trained with fine-grained supervision.

\begin{table}[t]
\centering
\caption{The performance of baseline algorithms and RL algorithms on three datasets when using MLP as the backbone model and ERR@10 as the reward function. The best method is marked in bold. "$\dagger$" indicates significantly worse than the best method at the $p < 0.05$ level using the two-tailed pairwise t-test.}
\renewcommand{\arraystretch}{1.2}
\label{tab:err_result}
\resizebox{0.99\columnwidth}{!}{
\begin{tabular}{c|c||cccc}
\toprule
Dataset & Algorithm & ERR@3 & NDCG@3 & ERR@10 & NDCG@10 \\ 
\midrule
\multirow{2}{*}{MSLR 30k}
 & PGRank & 0.3070 & 0.4119$^\dagger$ & 0.3487 & 0.4334$^\dagger$ \\
 & PPG & 0.3050$^\dagger$ & 0.4084$^\dagger$ & 0.3466$^\dagger$ & 0.4303$^\dagger$ \\
 & GRPO & \textbf{0.3081} & \textbf{0.4245} & \textbf{0.3506} & \textbf{0.4472} \\
 \midrule
\multirow{2}{*}{Istella-S}
 & PGRank & 0.6918$^\dagger$ & 0.6145$^\dagger$ & 0.7154$^\dagger$ & 0.6871$^\dagger$\\
 & PPG & 0.6906$^\dagger$ & 0.6136$^\dagger$ & 0.7147$^\dagger$ & 0.6869$^\dagger$ \\
 & GRPO & \textbf{0.7007} & \textbf{0.6291} & \textbf{0.7230} & \textbf{0.7079} \\ 
 \midrule
\multirow{2}{*}{Yahoo!}
 & PGRank & 0.4318 & 0.6929$^\dagger$  & 0.4683 & 0.7591$^\dagger$  \\
 & PPG & 0.4309$^\dagger$ & 0.6916$^\dagger$  & 0.4675$^\dagger$  & 0.7579$^\dagger$  \\
 & GRPO & \textbf{0.4334} & \textbf{0.7034} & \textbf{0.4694} & \textbf{0.7669} \\
 \bottomrule
\end{tabular}
}
\end{table}

Figure~\ref{fig:ndcg_curve} further illustrates the performance of different algorithms on the Yahoo! dataset using MLP as the backbone ranking model. 
The subfigure above shows the validation performance over training steps. 
We observe that the three supervised learning baselines (i.e., AttentionRank, CrossEntropy, and LambdaRank) converge quickly—typically within about 1,000 steps—but subsequently suffer from overfitting, as evidenced by a gradual decline in validation performance. 
In contrast, the coarse-grained RL methods (i.e., PGRank, PPG, and GRPO) take approximately 2,000 steps to reach convergence, but once converged, they exhibit stable performance without clear signs of overfitting.
On the other hand, the subfigure below presents the reward values recorded during training for the three RL-based methods. 
All three algorithms show a steady increase in reward over time. 
Among them, PGRank achieves slightly higher reward values than PPG and GRPO throughout training, while its validation performance does not significantly surpass that of the other two, suggesting that PGRank may be more prone to overfitting to the reward signal.

To explore the effectiveness and generalizability of employing coarse-grained relevance rewards for optimizing ranking models, we conduct additional experiments using the ERR@10 as the reward function.
The results are presented in Table~\ref{tab:err_result}.
We can observe that ranking models can also be effectively optimized using the ERR@10 relevance signal.
Furthermore, they demonstrate comparable performance when optimized with different relevance signals, although there are slight variations depending on the dataset and the chosen reward. 
Specifically, NDCG@10 tends to yield better results on the MSLR 30k and Istella-S datasets, whereas ERR@10 proves to be more effective on the Yahoo! dataset.
Therefore, in practical search or recommendation scenarios, one can flexibly choose appropriate reward functions or evaluation metrics for optimization based on his/her specific needs.

Overall, compared to fine-grained supervised learning methods, coarse-grained RL approaches typically converge more slowly.
This would not be a concern for large-scale search or recommendation platforms, where abundant and diverse user interaction data is readily available. 
However, the reliance on extensive user feedback may limit the applicability of such methods in scenarios where user interactions are scarce, such as in cold-start settings or for newly deployed systems.
On the other hand, it might indicate that training ranking models with coarse-grained rewards suffers less from overfitting in the long term, probably because the task is much more difficult than supervised learning and the model has larger freedom to explore in the parameter space.

\subsection{Results on Fairness (RQ2)}\label{subsec:click_and_fair}
\begin{figure}[t]
    \centering    
    \includegraphics[width=0.48\textwidth]{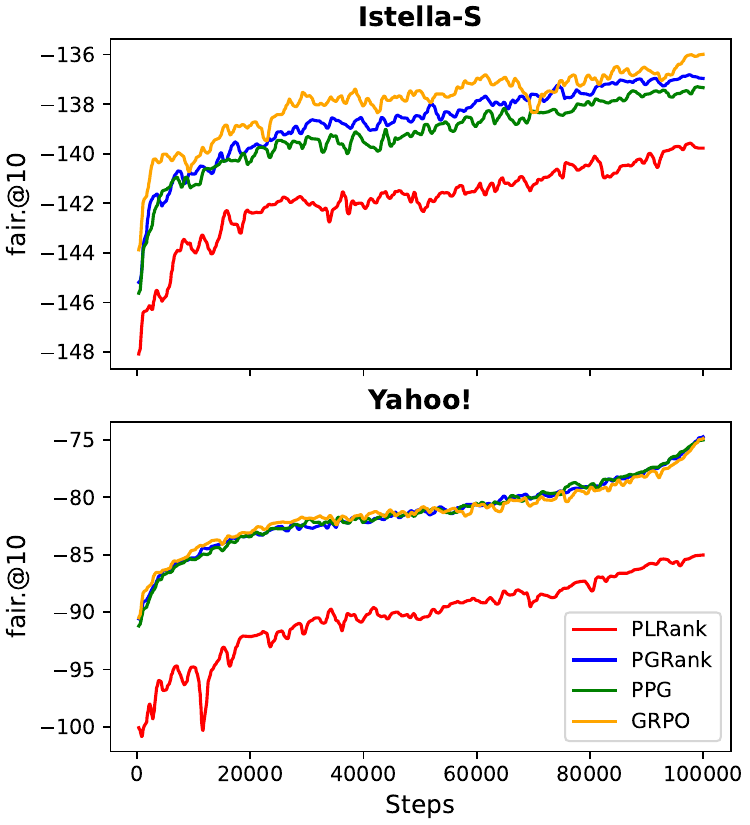}
    \caption{The fairness@10 of models trained using different RL algorithms based on the MLP backbone.}
    \label{fig:fair}
\end{figure}

To further investigate the effectiveness of RL-based methods in optimizing for coarse-grained signals with different characteristics, we additionally consider the more complex fairness ranking objective.
In contrast to traditional relevance-focused metrics, fairness-oriented metrics aim to promote equitable exposure across candidates.
Its complexity lies mainly in the following two points:
(1) non-decomposability, as fairness is typically defined over the distribution of exposure across the entire ranked list rather than individual items; and (2) conflicting optimization targets, since maximizing fairness may conflict with utility-oriented goals like NDCG.
These characteristics make fairness-aware reward signals challenging for conventional supervised learning methods, which struggle to optimize non-differentiable and list-level objectives. 
In contrast, RL provides a natural framework for handling such complex and structured feedback.
Here, we aim to evaluate the ability of coarse-grained RL algorithms to learn sophisticated ranking strategies.

Figure~\ref{fig:fair} illustrates the performance of different algorithms when optimizing ranking fairness. 
We observe that PLRank, which relies on fine-grained item-level relevance signals, performs significantly worse than the coarse-grained methods. 
On both datasets, PLRank exhibits slow convergence and converges to suboptimal performance levels; on the Yahoo! dataset, however, it exhibits high variance during training.
This degradation is likely due to the non-decomposable nature of fairness objectives mentioned above: fairness is inherently a list-level property, making it challenging for methods to assume item-wise factorization in the loss.
In contrast, the three coarse-grained RL methods consistently improve the fairness metric over time.
Specifically, GRPO shows a notable advantage on Istella-S, reaching a better final fairness score.
While on Yahoo!, all three methods are nearly indistinguishable, demonstrating high stability and robustness.
These results highlight the suitability of coarse-grained RL algorithms for optimizing complex, non-differentiable, and listwise objectives such as fairness. 
While supervised fine-grained methods struggle to capture the structural properties of fairness, coarse-grained RL methods adapt naturally to such reward formulations.

\begin{figure}[t]
    \centering
    \includegraphics[width=0.499\textwidth]{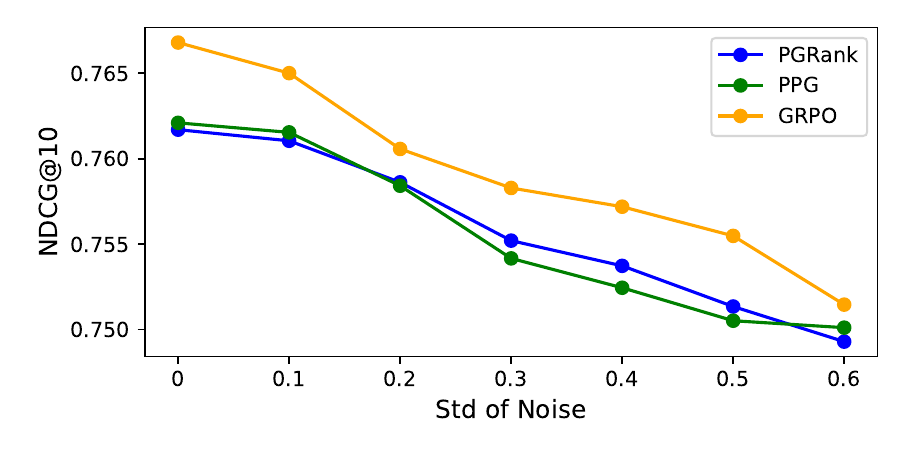}
    \caption{The influence of different levels of Gaussian reward noise on the results of coarse-grained RL methods.}
    \label{fig:robust}
\end{figure}

\begin{figure}[t]
    \centering
    \includegraphics[width=0.499\textwidth]{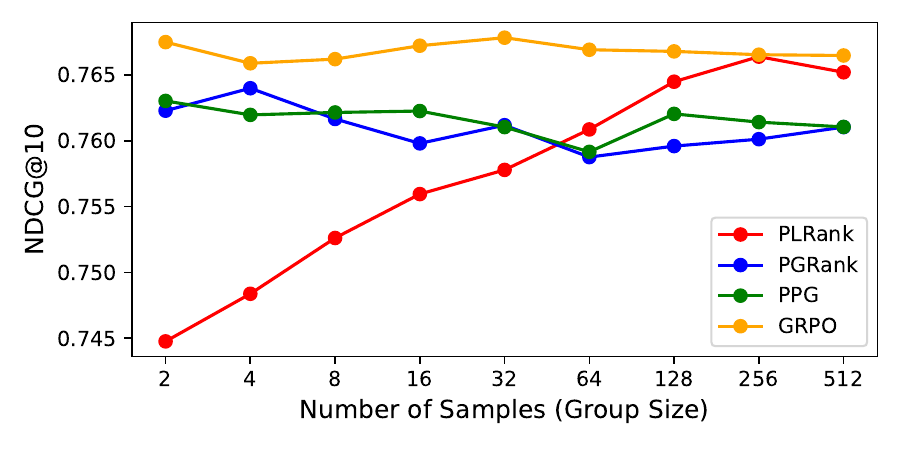}
    \caption{The influence of sampling count on the results of RL methods.}
    \label{fig:sample}
\end{figure}

\subsection{Analysis}
In this subsection, we analyze the characteristics of RL-based algorithms. 
Specifically, we investigate their robustness to noise in the reward signal, their sensitivity to the number of sampled ranking lists, and their behavior in multi-objective optimization settings. 
For simplicity, all experiments in this part are conducted on the Yahoo! dataset using the MLP as the backbone ranking model.

\subsubsection{Robustness against Noise (RQ3)}\hfill\newline
We first investigate the robustness of coarse-grained RL algorithms to noise in reward signals. 
In real-world scenarios, user feedback, such as engagement time or satisfaction, is often noisy, sparse, and difficult to model precisely. 
This stochasticity can stem from various factors, including user intent ambiguity, interface design, or random behavior.
As such, a ranking algorithm’s ability to withstand reward noise is critical for its practical effectiveness and application.
To simulate this condition, we introduce additive Gaussian noise to the reward signal of each sampled ranking list during training. 
Specifically, we perturb each reward with zero-mean Gaussian noise of varying standard deviations (std, or $\sigma$), thereby simulating different levels of noise, uncertainty, or inconsistency in user feedback.
Here, we aim to assess how well coarse-grained RL methods maintain their performance under increasing levels of noise. 
A desirable algorithm should degrade gracefully and remain stable even when the feedback signal is partially corrupted. 
By analyzing performance trends under controlled noise settings, we can gain deeper insight into each algorithm’s robustness and reliability in real-world environments.

Figure~\ref{fig:robust} illustrates the performance of three coarse-grained RL methods, PGRank, PPG, and GRPO, under varying levels of Gaussian noise injected into the reward signal. 
The x-axis denotes the standard deviation $\sigma$ of the zero-mean noise, while the y-axis shows NDCG@10 on the Yahoo! dataset.
We observe that, as the std of the injected noise increases, all methods experience a gradual decline in performance, confirming that noisy feedback introduces optimization challenges. 
However, the overall degradation is moderate and controlled: even under a relatively high noise level ($\sigma = 0.6$), all methods retain reasonable ranking quality (i.e., NDCG@10 > 0.75). 
This indicates that coarse-grained RL algorithms possess a degree of inherent robustness to stochasticity in reward signals, which is crucial for real-world deployment.
Besides, across all noise levels, GRPO consistently achieves the highest NDCG@10, demonstrating superior robustness compared to PGRank and PPG by a clear margin. 
It highlights that while noise in the reward signal does pose a challenge, coarse-grained RL methods—particularly GRPO—are resilient under such conditions. 
This reinforces the practical value of GRPO in settings where user feedback may be imperfect, inconsistent, or partially corrupted.

\subsubsection{Number of Samples (RQ4)}\hfill\newline
The second factor we investigate is the number of samples conducted on each query, which plays a critical role in RL-based ranking.
In our setting, the number of samples determines how many candidate ranking lists are drawn for each query. 
It directly affects both the estimation quality of the expected reward and the computational cost of training.
On the other hand, from a practical perspective, the sampling process can be seen as simulating how many times a real-world user might interact with the system under the same query. 
A robust algorithm should ideally perform well even with a small number of samples, since large-scale repeated interactions are often infeasible in real applications.
Therefore, in this experiment, we aim to assess the efficiency and robustness of RL-based methods under different levels of sampling. 
By varying the number of samples and observing performance trends, we evaluate whether these algorithms can achieve competitive results with minimal sampling, which is a critical property for their scalability and deployment in realistic retrieval environments.

\begin{figure}[t]
    \centering
    \includegraphics[width=0.499\textwidth]{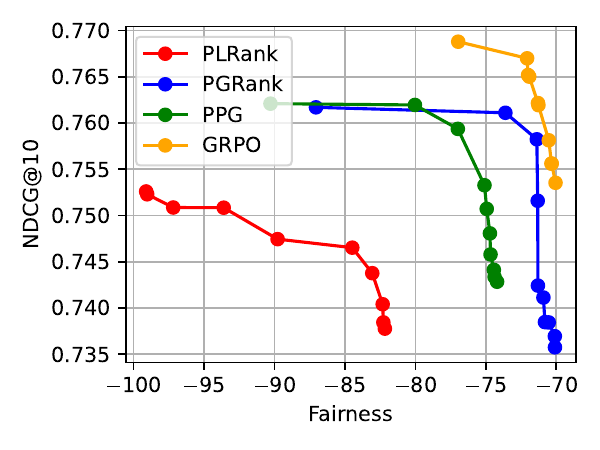}
    \caption{The fairness-NDCG trade-off of RL methods in multi-objective optimization.}
    \label{fig:multi}
\end{figure}

Figure~\ref{fig:sample}  illustrates how different algorithms perform under varying sampling sizes, measured by NDCG@10 on the Yahoo! dataset.
The three coarse-grained RL methods, PGRank, PPG, and GRPO, exhibit remarkable robustness to the number of samples. 
Even with as few as 2 sampled ranking lists per query, they can still achieve competitive performance, with only minor fluctuations as the sample count increases.
This indicates that these methods are highly sample-efficient, capable of learning effective ranking strategies even under minimal feedback scenarios.
Among them, GRPO consistently outperforms the others across all sampling settings. 
Notably, it shows the smallest sensitivity to sampling size, maintaining top-tier performance across the board. 
This stability highlights GRPO’s strong generalization ability and low-variance gradient estimation, making it particularly suitable for real-world environments where repeated user interactions per query are costly.
In contrast, fine-grained PLRank exhibits a stronger dependency on the number of samples compared to coarse-grained methods.
Its performance improves steadily with more samples, starting significantly below the RL-based methods at low sampling sizes, and approaching GRPO-level performance when the sampling size reaches 256 or more.
Therefore, these results confirm that coarse-grained RL methods, especially GRPO, offer superior sample efficiency and robustness, enabling effective and robust learning with minimal interaction cost. 
In contrast, fine-grained methods such as PLRank require substantially more samples to achieve comparable performance, limiting their practical applicability in resource-constrained environments.

\subsubsection{Multi-objective Training (RQ5)}\hfill\newline
The final setting we explore is multi-objective optimization.
In practical applications, stakeholders such as users, content providers, and platforms may each have different optimization goals. Thus, optimizing for a single objective (e.g., relevance) is rarely sufficient.
In this experiment, we consider a weighted combination of NDCG and fairness-based rewards, two metrics that often exhibit conflicting priorities, varying the trade-off coefficient to simulate different prioritization scenarios. 
This setup enables us to evaluate each algorithm’s ability to balance competing objectives and assess whether RL methods can adaptively navigate the trade-off frontier between different factors.
Therefore, this experiment serves as a testbed to analyze the flexibility and adaptability of coarse-grained RL methods, particularly their ability to discover effective policies when the reward landscape becomes more complex and multi-dimensional.

Figure~\ref{fig:multi} illustrates a clear trade-off between relevance and fairness in multi-objective optimization. 
It reveals that all coarse-grained RL methods substantially outperform the fine-grained PLRank in the multi-objective optimization of relevance and fairness. 
Specifically, the coarse-grained approaches consistently achieve higher NDCG scores across the entire fairness spectrum, while also demonstrating better fairness performance. 
This indicates that coarse-grained methods can deliver superior outcomes in both objectives simultaneously, rather than sacrificing one for the other.
On the other hand, among the coarse-grained methods, GRPO demonstrates clear superiority over PGRank and PPG. 
It not only achieves the highest NDCG values, but also has the best fairness performance, comparable to that of PGRank. 
More importantly, under any given fairness constraint, GRPO is able to maintain a higher level of relevance compared to the other methods. 
This implies that GRPO achieves Pareto dominance, offering a more favorable trade-off between relevance and fairness across the optimization spectrum.
Therefore, GRPO demonstrates stronger adaptability and generalization in multi-objective settings, making it more suitable for real environments with complex and diverse requirements.
\section{Conclusions}
In this paper, we investigate the potential of RL to optimize ranking models using coarse-grained reward signals, which are more accessible and better aligned with real-world user engagement than traditional fine-grained relevance labels.
We systematically compare various RL algorithms—including both classical policy gradient methods and the state-of-the-art GRPO—across multiple model architectures and ranking objectives.
Experimental results demonstrate that coarse-grained methods, particularly GRPO, can match but often surpass supervised learning baselines on standard relevance metrics, despite relying solely on listwise reward signals. 
Moreover, RL approaches exhibit notable advantages in optimizing complex and non-differentiable objectives such as ranking fairness. 
We also analyze their robustness to noisy feedback, efficiency under limited sampling, and their capability of balancing multiple objectives in multi-goal ranking tasks. 
It suggests a promising direction toward more scalable, flexible, and goal-aligned ranking systems, especially in scenarios where fine-grained labels are expensive, biased, or unavailable. 
Future work includes developing RL algorithms tailored to ranking scenarios and extending this framework to incorporate real-world user interactions in online environments.


\bibliographystyle{ACM-Reference-Format}
\bibliography{sample-base}

\end{document}